\documentclass[aps,pra,twocolumn,superscriptaddress,showpacs,floatfix]{revtex4}
\usepackage[dvipdfmx]{graphicx,color}
\usepackage{bm} 
\begin{document}

\title{Interaction modulation in a long-lived Bose-Einstein condensate by rf coupling}
\author{Kosuke Shibata}
\affiliation{Department of Physics, 
Gakushuin University Toshima, Tokyo 171-8588, Japan}
\email{shibata@qo.phys.gakushuin.ac.jp}
\author{Aki Torii}
\affiliation{Department of Physics, 
Gakushuin University Toshima, Tokyo 171-8588, Japan}
\author{Hitoshi Shibayama}
\affiliation{College of Industrial Technology, Nihon University, Narashino 275-8576,
Japan}
\author{Yujiro Eto}
\affiliation{National Institue of Advanced Industrial Science and Technology (AIST), NMIJ, Tsukuba 305-8568, Japan}
\author{Hiroki Saito}
\affiliation{Department of Engineering Science, University of Electro-Communications, Chofu, Tokyo 182-8585, Japan}
\author{Takuya Hirano}
\affiliation{Department of Physics, 
Gakushuin University Toshima, Tokyo 171-8588, Japan}

\date{\today}

\begin{abstract}
We demonstrate modulation of the effective interaction between the magnetic sublevels of the hyperfine spin $F=1$ 
in a $^{87}$Rb Bose-Einstein condensate by Rabi coupling with radio-frequency (rf) field.
The use of the $F=1$ manifold enables us to observe the long-term evolution of the system owing to the absence of inelastic collisional losses.
We observe that the evolution of the density distribution reflects the change in the effective interaction between atoms due to rf coupling. 
We also realize a miscibility-to-immiscibility transition in the magnetic sublevels of $m = \pm 1$ by quenching the rf field.
Rf-induced interaction modulation in long-lived states as demonstrated here
will facilitate the study of out-of-equilibrium quantum systems.
\end{abstract}

\maketitle
\section{Introduction}
Nonequilibrium dynamics are ubiquitous across wide areas ranging  from the early universe to condensed matter physics.
Phase transition dynamics involves highly nonequilibrium phenomena.
Crossing the critical point of the phase transition, where the characteristic time scale diverges, gives rise to nonequilibrium. 
Understanding nonequilibrium phenomena in those situations has been fundamentally important. 

A cold atom system offers a good platform for studying nonequilibrium dynamics.
Systems of this type allow a clear comparison between experiment and theory owing to their high controllability.
Furthermore, the dynamics in a cold atom system is usually slow enough to be observed with practical time resolution.
Quantized vortices \cite{Weiler_Spontaneous_2008} and solitons \cite{Lamporesi_Spontaneous_2013} 
were created in phase transitions induced by thermally quenching an atomic gas to a Bose-Einstein condensate (BEC).
Recent studies \cite{Lamporesi_Spontaneous_2013,Navon_critical_2015} confirmed
the power-law dependence of defect number on quench time, predicted by the Kibble-Zurek (KZ) theory \cite{Kibble_Topology_1976, Zurek1985}.
We can also explore the quantum phase transitions \cite{Eisert_Quantum_2015} and quantum KZ theory \cite{Zurek_QPT_2005} with cold atoms.
The polar to broken-axisymmetry quantum phase transition in a spin-1 BEC \cite{Sadler_Spontaneous_2006} 
is a  promising candidate for testing quantum KZ theory \cite{Lamacraft_Quantum_2007, Saito_Topological_2007}.
The power law in the spin excitation during the broken-axisymmetry phase transition 
was recently shown to be in good agreement with quantum KZ theory \cite{Anquez_Quantum_2016}.
The dynamics of the Mott-superfluid quantum phase transition of atoms in an optical lattice 
was shown to be complex beyond power-law scaling \cite{Braun_Emergence_2015}.

The miscible-immiscible phase transition can occur in a two-component BEC
depending on the interaction between atoms.
A test of quantum KZ theory with an engineered miscible-immiscible transition has been proposed \cite{Sabbatini_Phase_2011,Sabbatini_Causality_2012}. Such engineered transitions can be achieved with 
optical Raman dressing of the atomic states \cite{Lin_Spin_2011} or Rabi coupling \cite{Nicklas_Rabi_2011}.
Although the proposed test has not been realized, 
scaling of the spin-spin correlations was observed during the short-term evolution after a sudden quench of the coupling \cite{Nicklas_Observation_2015}.

In the present paper, we demonstrate rf-induced modulation of the effective interaction 
between different magnetic sublevels in the lowest hyperfine state with the total spin $F=1$ in a $^{87}$Rb BEC.
In the previous studies, 
interaction modulation via Rabi coupling has been achieved in the pair of $|F, m \rangle = | 1, -1 \rangle $ and $|2, 1 \rangle$ states of a $^{87}$Rb BEC \cite{Nicklas_Rabi_2011,Nicklas_Nonlinear_2015,Nicklas_Observation_2015},
which inevitably suffer from inelastic collisional losses.
Using the $F=1$ states with a small loss rate, we can study the long-term effect of the interaction modulation on nonequilibrium dynamics.
We couple the pair of $\vert 1, -1 \rangle$ and $\vert 1, 0 \rangle$ states
and the pair of $\vert 1, -1 \rangle$ and $\vert 1, +1 \rangle$ states, 
which are miscible and immiscible pairs, respectively,
according to the known values of the $s$-wave scattering lengths \cite{Kempen_Interisotope_2002}.
For both of these pairs, we observe that the dynamics are affected by the modulation of the effective interaction due to rf coupling.
Furthermore, we demonstrate the miscibility-to-immiscibility transition in the pair of $\vert 1, -1 \rangle$ and $\vert 1, +1 \rangle$ states by quenching the rf coupling.

The paper is organized as follows. 
We describe the experimental setup in Sec. \ref{sec: setting}.
In Sec. \ref{sec: Result}, we present the results of experiments and numerical simulation.
We discuss the applications of the demonstrated interaction modulation in Sec. \ref{sec: Discussion}.
We conclude in Sec. \ref{sec: conclusion}.

\section{Experimental Setup}\label{sec: setting}
We prepare a BEC of $^{87}$Rb in an optical trap elongated along the horizontal axis ($x$-axis) 
by transferring a BEC in the $\vert 2,+2 \rangle$ state produced in a magnetic trap.
The optical trap is formed by two horizontal beams with wavelengths of $980$ nm and $1064$ nm
intersecting at a right angle.
The axial and radial trapping frequencies of the optical trap are $2\pi \times (28, 180)$ Hz, respectively. 
When the magnetic trap is turned off for the transfer,  
the spin is flipped, leaving a BEC in the $\vert 2,-2 \rangle$ state in the optical trap.
After switching off the magnetic trap, 
we wait 500 ms for the magnetic field to settle
and transfer the BEC of typically $2 \times 10^5$ atoms to the $\vert 1,-1 \rangle$ state by a microwave $\pi$ pulse with a frequency of 6.818770 GHz, 
corresponding to a magnetic field of 7.575 Gauss.

\begin{figure}
 \centering
 \includegraphics[width=8.6cm]{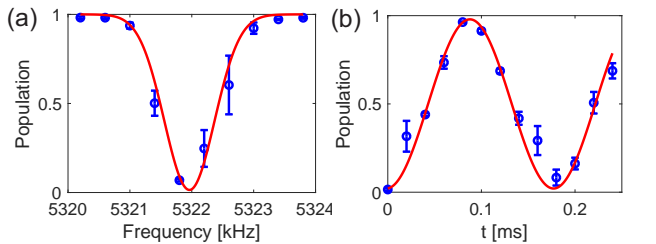}
 \caption{(Color online) Performance of the two-photon rf transition between the $|1,-1\rangle$ and $|1,+1\rangle$ states.
(a) Typical two-photon rf spectrum. The population of the $|1,-1\rangle$ state after applying a Gaussian rf pulse is plotted. 
The solid line is a Gaussian fit.
(b) Observed Rabi oscillation. 
The population of the $|1,+1\rangle$ state is plotted against the rf pulse width $t$.
The solid curve is a sinusoidal fit.}
\label{fig: spec} 
\end{figure}

The stability in the bias magnetic field is important
to keep the rf field resonant to the transition between the magnetic sublevels.
The experimental room is surrounded by magnetic shielding walls to suppress outside noise.
The coil generating the bias field is driven by a stable current source (ILX Lightwave, LDX-3232).
In addition, a magnetic field fluctuation synchronized with the power line at 50 Hz is detected by spin-echo ac magnetometry \cite{Eto_Spin_2013} using the $|1,-1\rangle$ and $|1,0\rangle$ states and 
we apply a canceling magnetic field at 50 Hz to bring the ac field below 0.1 mG.

The rf frequency is set on the basis of precise spectroscopy between the magnetic sublevels in the $F=1$ state.
The magnetic field is stable enough to ensure sub-kHz precision.
A typical spectrum of the two-photon rf transition between
$| 1,-1 \rangle$ and $|1, +1 \rangle$ states is shown in Fig. \ref{fig: spec}(a).
The population of the $|1, -1 \rangle$ state after a Gaussian rf pulse 
with a pulse width (standard deviation) of 310 $\mu$s is plotted. 
The data is fitted by a Gaussian $\propto \exp \left[-(f-f_0)^2/(2\sigma^2) \right]$,
yielding $\sigma = 406(2)$ Hz and $f_0 = 5321.96(4)$ kHz. 
The two-photon Rabi oscillation at the resonance frequency $f_0$ is shown in Fig. \ref{fig: spec}(b).
The effective Rabi frequency in this case is estimated to be $\Omega = 2\pi \times 5.6(1)$ kHz
by fitting a sinusoidal function to the data.

We study the dynamics of pairs of magnetic sublevels coupled by an rf field.
We use a waveform generator (Keysight Technologies Inc., 33611A) to start applying an rf wave 100 ms after the preparation of the $|1,-1 \rangle$ state.
The atoms are held in the optical trap under the rf irradiation for a variable holding time $T$.
When we couple the $|1,-1 \rangle$ and $|1, +1 \rangle$ states,
we use an rf wave envelope with falling and rising edges smoothed to suppress an undesired transition to the intermediate $|1,0\rangle$ state.
For comparison, we also study the dynamics without rf coupling.
In the experiments without rf coupling,
a $\pi/2$ rf pulse is applied to prepare a mixture of two sublevels and then the atoms are held in the absence of an rf field for $T$.
After that hold period, we release the atoms from the trap and take an absorption image with a time-of-flight (TOF) of 17.5 ms.
The magnetic sublevels are separated by an applied magnetic field gradient during the time-of-flight.

\section{Results}\label{sec: Result} 
\subsection{Overview}\label{subsec: Overview}

\begin{figure}
 \centering
 \includegraphics[width=8.6cm]{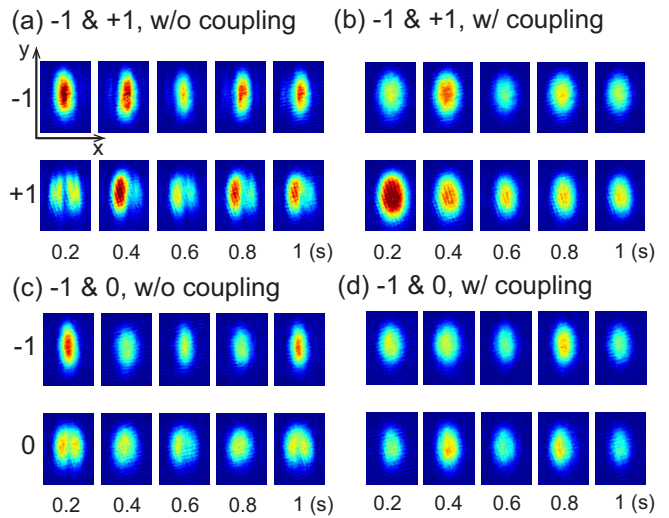}
 \caption{(Color online) Typical TOF images for holding times of $T= 0.2, 0.4, \cdots, 1$ s.
(a), (b) $|1,-1\rangle$ and $|1,+1\rangle$ states without and with rf coupling, respectively.
(c), (d) $|1,-1\rangle$ and $|1, 0\rangle$ states without and with rf coupling, respectively.
The $|1,-1\rangle$ state is displayed in the top row.
The horizontal ($x$) and vertical ($y$) dimensions of each image are $150$ $\mu$m and 
$210$ $\mu$m, respectively.
}
 \label{fig: TOFimages} 
\end{figure}

We show typical TOF images for holding times of $T= 200, 400, \cdots, 1000$ ms
with and without coupling in Fig. \ref{fig: TOFimages}.
The Rabi frequencies for the $|1, -1 \rangle$ - $|1,+1 \rangle$ pair and the $|1, -1 \rangle$ - $|1, 0 \rangle$ pair
are measured to be $2\pi \times 5.6(1) $ kHz and $2\pi \times 2.557(8)$ kHz, respectively. 
We can see that the evolution of the density distribution for each pair is changed by rf coupling.
These changes can be attributed to
the modulation of the effective interaction between atoms induced by rf coupling \cite{Nicklas_Rabi_2011}.
Whereas previous rf coupling experiments have exploited the $|1, 1 \rangle$ and $|2, -1 \rangle$ states with the same linear Zeeman shift
insensitive to magnetic field fluctuation
\cite{Nicklas_Rabi_2011,Nicklas_Nonlinear_2015,Nicklas_Observation_2015},
we study the dynamics of magnetically sensitive pairs of $F=1$ states 
owing to the stable magnetic environment.
These lowest hyperfine states do not suffer from inelastic losses
and their long-term evolution can be observed.

The miscibility of the two states is evaluated by the miscibility parameter 
\begin{equation}
\Delta = \frac{a_{ii}a_{jj}}{a_{ij}^2},
\end{equation}
where $a_{ii}$, $a_{jj}$ and $a_{ij}$ are the $s$-wave scattering lengths between the states labeled $i$ and $j$.
The two states are miscible when $\Delta >1$ and immiscible when $\Delta<1$ in a homogeneous system.
When the two states are coupled with a sufficiently high Rabi frequency,
the system is suitably described by the dressed states, with the effective scattering lengths given by \cite{Jenkins_Dynamic_2003}
\begin{equation}\label{Eq: a++}
a_{+,+} = a_{-,-} = \frac{1}{4}(a_{ii}+2a_{ij}+a_{jj})
\end{equation} 
and
\begin{equation}\label{Eq: a+-}
a_{+,-} = \frac{1}{2}(a_{ii}+a_{jj}),
\end{equation} 
where 
$| + \rangle =\frac{1}{\sqrt{2}} (|i \rangle + |j \rangle)$ and 
$| - \rangle =\frac{1}{\sqrt{2}} (|i \rangle - |j \rangle)$ denote the dressed states.
Using Eqs. (2) and (3), the miscibility condition of the $|\pm\rangle$ states, 
$a_{++} a_{--} / a_{+-}^2 > 1$, can be written as $a_{ii} + a_{jj} > 2 a_{ij}$.
Effective modulation of the scattering lengths thus leads to the reversal of miscibility for the pairs of magnetic sublevels of $^{87}$Rb \cite{Nicklas_Rabi_2011,Sinatra_Binary_2000,Jenkins_Dynamic_2003}.
In our experiments, the initial state under rf coupling is 
$| i \rangle = \frac{1}{\sqrt{2}} (|+\rangle + |-\rangle)$. 
When a phase separation occurs between the $|\pm\rangle$ states, 
the density distributions of the $|i \rangle$ and $|j\rangle$ states are also modulated.

The bare $|1, -1 \rangle$ and $|1,+1\rangle$ states are predicted to be immiscible.
The scattering lengths in the bare $|1, -1 \rangle$ and $|1,+1\rangle$ states are 
$(a_{-1,-1}, a_{-1,+1}, a_{+1,+1}) = (100.40, 101.32, 100.40) a_\mathrm{B}$ with 
$a_\mathrm{B}$ being the Bohr radius \cite{Kempen_Interisotope_2002} 
and $\Delta <1$. 
We expect the rf-coupled dressed states to be miscible,
because the coupling strength $\hbar\Omega$ is much larger than 
the characteristic energy of miscibility $\sqrt{g_{ij}^2 - g_{ii} g_{jj}} \rho \sim \hbar \times 10$ Hz, where $\rho$ is the atom density.
The observed distributions are consistent with 
the assumption that the two states are miscible.
In contrast, the miscible parameter of the $|1, -1 \rangle$ and $|1, 0\rangle$ pair is $\Delta = 1.005 >1$,
on the basis of the scattering lengths $(a_{-1,-1}, a_{-1,0}, a_{0,0}) = (100.40, 100.40, 100.86) 
a_\mathrm{B}$\cite{Kempen_Interisotope_2002}. 
However, a weak splitting of the $|1, 0\rangle$ distribution is observed 
in the no-coupling case and the splitting disappears in the coupling case.
This behavior, seemingly contradictory to the above argument on miscibility, can be ascribed to the fact that the scattering lengths are comparable, as we shall discuss later.

\subsection{Dynamics of the $|1,-1\rangle$ and $|1,+1\rangle$ pair}

\begin{figure}
 \centering
 \includegraphics[width=8.6cm]{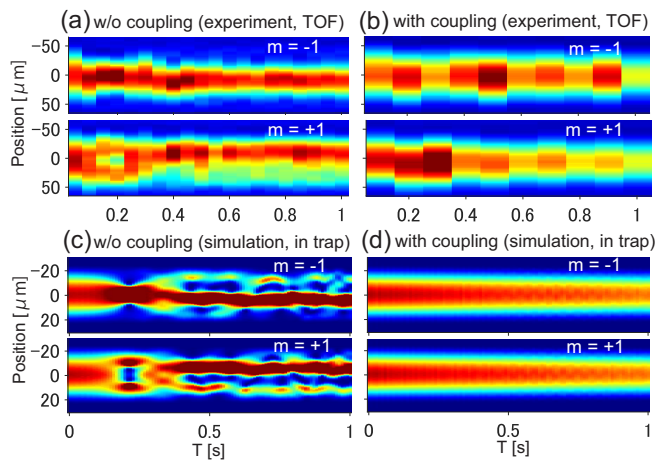}
 \caption{(Color online) Time evolution of the radially integrated densities of the $|1,-1\rangle$ and $|1,+1\rangle$ states
without [(a), (c)] and with [(b), (d)] rf coupling.
(a) and (b) are experimental results, while (c) and (d) are numerical simulation results. 
Results averaged over data from several experiments are shown in (a) and (b).
Since the population imbalance varies owing to the Rabi oscillation,
the data with population imbalance within 1:3 and 3:1 are averaged in (b)
and the data with 1:1 are used in (d).
}
 \label{fig: 1and-1dynamics}
\end{figure}

We study the evolution of the radially integrated densities 
$\tilde{n}_{m}(x) = \int n_{m} (\bm{r}) dydz$
with $n_{m}$ being the density of the magnetic sublevel $m$.
The experimental results without and with rf coupling are shown in Figs. \ref{fig: 1and-1dynamics}(a) and \ref{fig: 1and-1dynamics}(b), respectively.
Without coupling, the two components are dynamically separated along the $x$-axis.
Symmetric separation appears in the $| 1, -1 \rangle$ state around $T=$ 200 ms,
and the distribution becomes asymmetric for longer holding times.
When the rf field is applied, we observe no separation.

\begin{figure}
 \centering
 \includegraphics[width=8.6cm]{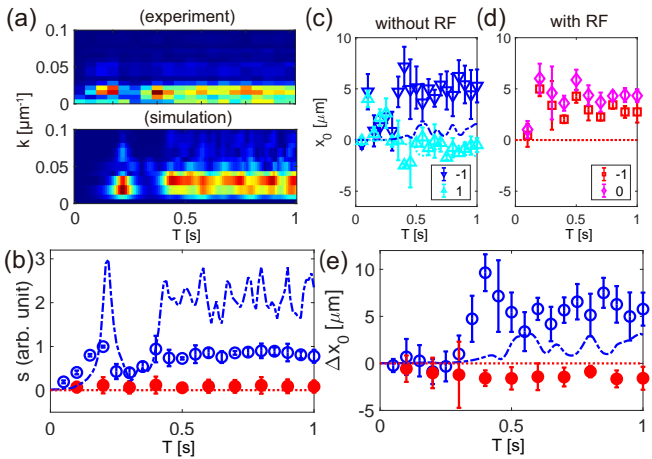}
 \caption{(Color online) Time evolution of the characteristic quantities of the dynamics of the $|1,-1 \rangle$ and $|1, +1 \rangle$ states.
(a) The Fourier-space distribution of the density difference $|\hat n(k)|$
of experiment (top) and numerical simulation (bottom).
(b) The separation parameter $s$.
(c), (d) The center of mass $x_0$ without (c) and with (d) coupling.
The simulation results are indicated by the dotted and dot-dashed lines.
(e) The relative position $\Delta x_0$. 
In (b) and (e), the open circles and dot-dashed line (closed circles and dotted line) 
represent the experimental and numerical results without (with) coupling, respectively.
} 
 \label{fig: 1and-1analysis}
\end{figure}

To extract the characteristic features of the dynamics,
we calculate the Fourier transform of the difference between radially integrated densities of the two components,
given by $\hat{n}(k) = \mathcal{F}\left[\tilde{n}_{1}(x) - \tilde{n}_{-1}(x)\right]$, 
where $k$ is the wavenumber in the $x$ direction.
The distribution of $|\hat{n}(k)|$ for the no-coupling case is shown in Fig. \ref{fig: 1and-1analysis}(a).
We evaluate the degree of separation with the first-order moment in the Fourier space defined by
\begin{equation}
s = \int_{k>0} k |\hat{n}(k)| dk,
\end{equation}
plotted against $T$ in Fig. \ref{fig: 1and-1analysis}(b).
Note that $s$ increases when phase separation occurs and 
larger wavenumber components of the density appear.
The peak of $s$ around $T=200$ ms corresponds to the splitting of the $|1, +1 \rangle$ state.

We also investigate the effect of coupling on the center-of-mass motion.
The centers of mass of each component, $x_0^{(\pm1)}$, without and with coupling are plotted in Figs. \ref{fig: 1and-1analysis}(c) and \ref{fig: 1and-1analysis}(d), respectively.
The center of each component at $T=0$ is taken to be zero.
The relative center of mass of the two components defined by $\Delta x_0 \equiv x_0^{(-1)} - x_0^{(+1)}$ is plotted in Fig. \ref{fig: 1and-1analysis}(e).
We observe that the two components remain overlapped $(\Delta x_0 \simeq 0)$
when rf coupling is introduced.

We compare our experimental results with a numerical simulation based on the Gross-Pitaevskii equations given by
\begin{eqnarray}
i\hbar\frac{\partial \psi_1}{\partial t} &=&\left( -\frac{\hbar^2}{2M}\nabla^2 +V_1 \right) \psi_1 + \frac{\hbar\Omega}{2} \psi_2 + g_{11} |\psi_1|^2\psi_1  \nonumber \\
&&+ g_{12} |\psi_2|^2\psi_1 -i \hbar \Gamma \psi_1,
\end{eqnarray}
\begin{eqnarray}
i\hbar\frac{\partial \psi_2}{\partial t} &=&\left( -\frac{\hbar^2}{2M}\nabla^2 +V_2 \right) \psi_2 + \frac{\hbar\Omega}{2} \psi_1 + g_{22} |\psi_2|^2\psi_2  \nonumber \\
&&+ g_{12} |\psi_1|^2\psi_2-i \hbar \Gamma \psi_2,
\end{eqnarray}
where $\psi_i$ is the macroscopic wave function, 
$M$ is the atomic mass,
$V_i$ is the trap potential, 
$g_{ij} = \frac{4\pi\hbar^2 a_{ij}}{M}$, 
and $\Omega$ is the Rabi frequency. 
The initial atom number in the simulation is $1.64 \times 10^5$ and the atom loss corresponding to the experiment is taken to be $\Gamma = 0.14$ s$^{-1}$.
The density distributions in the trap determined by the simulations 
without and with rf coupling are shown 
in Figs. \ref{fig: 1and-1dynamics}(c) and \ref{fig: 1and-1dynamics}(d), respectively.
The values of $s$, $x_0$, and $\Delta x_0$ obtained by simulation
are represented by the dotted and dot-dashed lines in Fig. \ref{fig: 1and-1analysis}.

The experimental results suggest that the experimental trapping potential 
exhibits a slight state dependence 
owing, for example, to the inhomogeneous residual magnetic field or the vector AC Stark shift induced by an elliptically polarized light field.
The splitting of the $\vert 1, 1 \rangle$ state, which we reproducibly observe in the experiment,
would not occur if the potential were perfectly symmetric, 
because $a_{-1,-1} = a_{+1,+1}$ and the roles of the $\vert 1, +1 \rangle$ and $\vert 1, -1 \rangle$ states should be the same.
A very small difference between $V_1$ and $V_2$ is introduced in the simulation
by setting $V_1 = 1.0005 V_2$.

The simulation reproduces the main features of the experiment.
The separation parameter $s$ has a peak around $T$ = 200 ms and remains roughly constant after $T$ = 400 ms.
The relative position $\Delta x_0$ remains zero in the case of coupling, while $\Delta x_0$ becomes nonzero in the absence of coupling.
However, the movement of atoms in the case of coupling is observed only in the experiment [see Fig. \ref{fig: 1and-1analysis}(d)].
Although the reason for this movement remains unclear, 
the kinetic energy of the atoms acquired during the state preparation might be responsible.
Even in the presence of such kinetic noises due to experimental imperfections,
the overlap of the dressed states remains robust.
The relative center of mass of the two coupled components is kept close to zero as shown in Fig. \ref{fig: 1and-1analysis}(e),
while both components move by several $\mu$m.
The stable overlap of up to 1 s observed in this experiment may be advantageous for various applications including spin squeezing \cite{Jenkins_Spin_2002}.
We note that the stable overlap implies a homogeneous coupling,
because an inhomogeneous coupling gives rise to spatial spin structures \cite{Matthews_Untwist_1999,Nicklas_Rabi_2011,Hamner_Phase}.

\subsection{Dynamics of the $|1,-1\rangle$ and $|1,0\rangle$ pair}
\begin{figure}
 \centering
 \includegraphics[width=8.6cm]{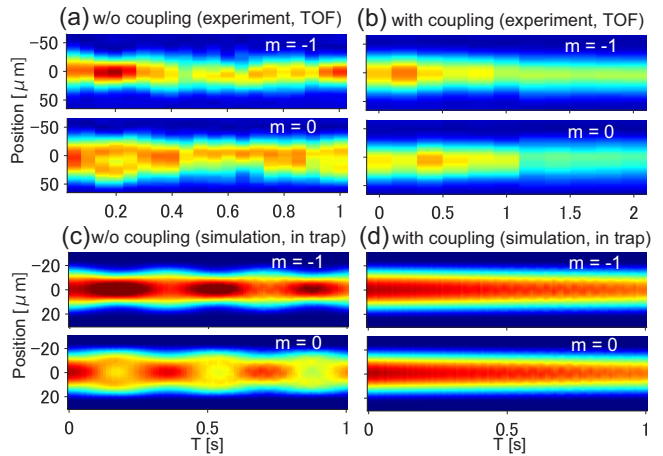}
 \caption{Time evolution of the radially integrated densities of the $|1,-1\rangle$ and $|1,0\rangle$ states.
without [(a), (c)] and with [(b), (d)] rf coupling.
(a), (b) are experimental results and (c), (d) are numerical simulation results.
Results averaged over data from several experiments are shown in (a) and (b).
The data with population imbalance within 4:6 and 6:4 are averaged in (b)
and the data with 1:1 are used in (d).
}
 \label{fig: -1and0dynamics}
\end{figure}

\begin{figure}
 \centering
 \includegraphics[width=8.6cm]{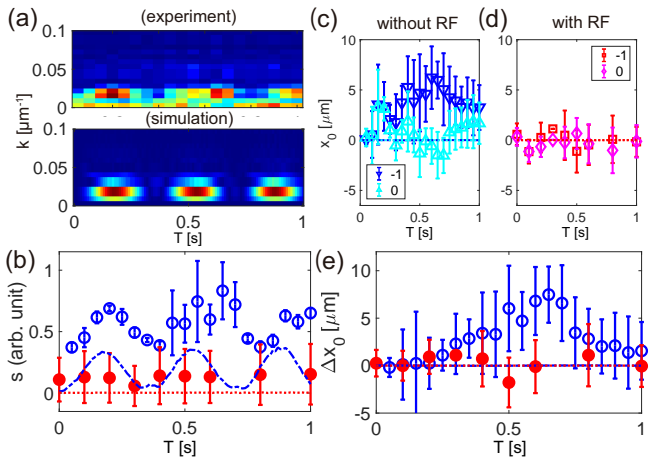}
 \caption{(Color online) The evolution of the characteristic quantities of the dynamics of the $|1,-1 \rangle$ and $|1, 0 \rangle$ states.
(a) The Fourier-space distribution of the density difference between the experimental (top) and numerical (bottom) results.
(b) The separation parameter $s$
of experimental and numerical results without (with) coupling.
(c), (d) The center of mass $x_0$ without (c) and with (d) coupling.
The numerical results are shown as dotted and dot-dashed lines.
(e) The relative position $\Delta x_0$. 
In (b) and (e), the open circles and dot-dashed line (closed circles and dotted line) represents the experimental and numerical results without (with) coupling, respectively.
}
 \label{fig: -1and0analysis}
\end{figure}

We show the evolution of the radially integrated densities of the $|1, -1 \rangle$ and $|1, 0 \rangle$ states 
in experiments without and with coupling in Figs. \ref{fig: -1and0dynamics}(a) and \ref{fig: -1and0dynamics}(b), respectively.
Demixing dynamics are observed in the bare states.
This seems incompatible with the miscibility condition, since 
the bare $|1,-1\rangle$ and $|1,0\rangle$ states are miscible $(\Delta > 1)$,
as mentioned in Sec. \ref{subsec: Overview}. 
The demixing of the two components stems from the difference in the interaction: $a_{0,0}>a_{-1,-1}$.
The atoms are initially prepared in the $|1,-1\rangle$ state, and the $\pi/2$ pulse generates a mixture of $|1, -1\rangle$ and $|1, 0\rangle$ states.
As a result, the $|1,0 \rangle$ state with larger interaction is pushed outward \cite{Timmermans_Phase_1998},
which leads to oscillatory behavior.
Oscillatory behavior can be seen in the Fourier components shown in Figs. \ref{fig: -1and0analysis}(a) and \ref{fig: -1and0analysis}(b).
In the coupled case,
we observe no separation for up to $T=2$ s as shown in Fig. \ref{fig: -1and0dynamics}(b),
whereas phase separation is expected for the strongly coupled miscible $(\Delta>1)$ states \cite{Nicklas_Rabi_2011}.
This is because the differences in scattering lengths are small. Substituting the bare scattering lengths into Eqs. (\ref{Eq: a++}) and (\ref{Eq: a+-}),
we obtain $a_{+,+} = a_{-,-} = 100.52 a_\mathrm{B}$, $a_{+,-} = 100.63 a_\mathrm{B}$, and the miscibility parameter $a_{+,+}a_{-,-}/a_{+,-}^2 =0.998$.
This indicates that the dressed states are only weakly immiscible and the instability dynamics are too slow to be observed within $T=2$ s.

We show the results of the numerical simulations in 
Figs. \ref{fig: -1and0dynamics}(c), \ref{fig: -1and0dynamics}(d) and \ref{fig: -1and0analysis}.
These numerical results are mostly consistent with the experimental ones, supporting the above discussion.
The slow oscillation of the atom centers in the no-coupling case is observed only experimentally [see Figs. \ref{fig: -1and0analysis}(c) and  \ref{fig: -1and0analysis}(e)].
Again, this is likely due to experimental imperfections such as a state-dependent or asymmetric potential.
When the coupling is applied, the center of each component remains around zero.

\subsection{Quench of coupling}

\begin{figure}
 \centering
 \includegraphics[width=8.5cm]{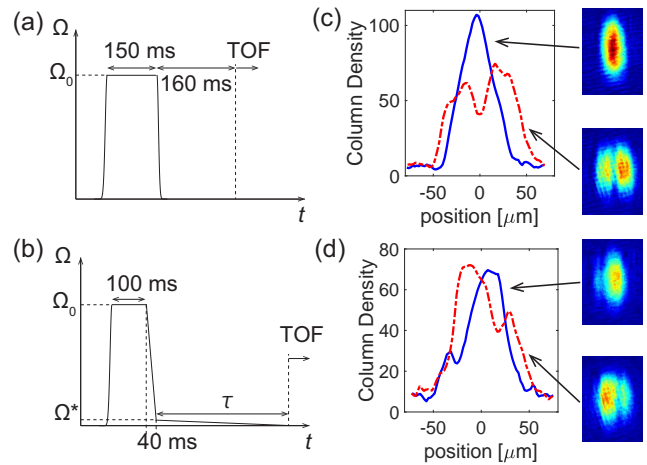}
 \caption{(Color Online) Quench experiment. 
             (a) [(b)] Time sequence for the fast [slow] quench.
             (c) [(d)] Typical TOF images and radially integrated densities after the fast [slow] quench.
                        The solid blue and red dot-dashed lines represent 
the column densities of the $|1,-1\rangle$ and $|1,+1\rangle$ states, respectively.
The data with population imbalance $\simeq 1:1$ are chosen in (c) and (d).
}
 \label{fig: Quench}
\end{figure}

We demonstrate the miscible-immiscible transition in the $|1,-1\rangle$ and $|1,+1\rangle$ pair by quenching the coupling strength.
We first describe the fast quench.
The time sequence of the fast quench experiment is depicted in Fig. \ref{fig: Quench}(a).
The two states initially undergo Rabi oscillation by a coupling of $\Omega_0 =2\pi \times 5.6$ kHz 
and then the coupling is instantaneously quenched.
The atoms are held in the trap for 160 ms after the quench and imaged.
A typical TOF image and the radially integrated column density are shown in Fig. \ref{fig: Quench}(c).
The density distribution is essentially the same as that observed in the evolution after the $\pi/2$ pulse (Fig. \ref{fig: 1and-1dynamics}(a)).
This is not surprising because, like the $\pi/2$ pulse, the fast quench induces an instantaneous change in the interaction energy.
We also perform slow quench experiments with the time sequence shown in Fig. \ref{fig: Quench}(b).
The initial coupling strength $\Omega_0$ decreases linearly to $\Omega^{*} = \Omega_0/15$ in 40 ms
and subsequently ramps down linearly to zero in $\tau$.
In our setup, $\tau$ can be as long as 2 s. The limitation comes from the memory size of the waveform generator (64 MSa).
A typical result after the slow quench with $\tau = 400$ ms is shown in Fig. \ref{fig: Quench}(d).
We observe a splitting in the $|1,-1\rangle$ state,
which is not observed in Figs. \ref{fig: 1and-1dynamics} and \ref{fig: Quench}(a).
This result suggests that the dynamics with the slow quench is qualitatively different
from the dynamics after the sudden state transfer \cite{Eto_Nonequilibrium_2016}.
We also find that the distribution after the slow quench is not deterministic.
For example, a pattern having more stripes is sometimes observed 
in a shot with the population imbalance almost equal to that in Fig. \ref{fig: Quench}(b). 
This implies that the dynamics with the slow quench is sensitive to subtle changes in experimental conditions.

The dynamics with the slow quench of the Rabi coupling has been little investigated experimentally or theoretically.
Although the time evolution following a sudden coupling quench was examined  \cite{Nicklas_Observation_2015},
the dynamics associated with a non-instantaneous coupling quench remain unclear.
Naively,  when the coupling is sufficiently decreased,
the original interaction will govern the evolution of the two components.
The dynamical behavior, however, cannot be understood so simply. 
Bogoliubov analysis might help understanding the dynamics.
The excitation spectra of a homogeneous BEC for general coupling strengths have been theoretically studied \cite{Tommasini2003}
and the predicted results roughly agree with the results of the sudden quench experiment \cite{Nicklas_Observation_2015}.
These theoretical results, however, cannot be applied to the dynamics of a trapped gas with slow quench.
In addition, the oscillation modes of the condensate may be excited during the non-instantaneous quench.
During a Rabi oscillation, the interaction energy oscillates as $\propto \frac{g_{11}}{2} \cos^4 \frac{\Omega}{2} t + \frac{g_{22}}{2} \sin^4 \frac{\Omega}{2} t + g_{12} \sin^2 \frac{\Omega}{2} t \cos^2 \frac{\Omega}{2} t$. 
Therefore, when the Rabi frequency matches a mode frequency,
the condensate will be resonantly excited.

\begin{figure}
 \centering
 \includegraphics[width=8.5cm]{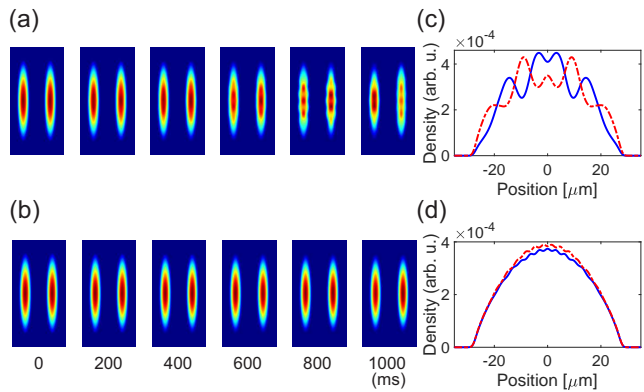}
 \caption{(Color Online) Numerically obtained density distributions for a
weak constant coupling.
             (a) [(b)] The density distributions of the $|1,-1\rangle$ (left) and $|1,+1\rangle$ (right) states with the coupling $\Omega_1$ [$\Omega_2$] at $t = 0, 200, ... ,1000$ ms.
            (c) [(d)] The density distribution along the condensate center axis at $t = 1000$ ms in (a) [(b)]. 
The $|1,-1\rangle$ and $|1,+1\rangle$ states are represented by the solid and dot-dashed lines, respectively.
}
 \label{fig: QuenchSim}
\end{figure}

To see this excitation,
we performed numerical simulations for a weak continuous coupling in Fig. \ref{fig: QuenchSim}.
The atoms are initially prepared in the $|1,-1 \rangle$ state and subsequently a constant coupling is applied continuously. When the coupling strength is  $\Omega_1= 2\pi \times 920/15$ (= 61.3) Hz,
the density modulation gradually starts to appear as shown in Fig. \ref{fig: QuenchSim}(a).
We note that the wavelength of the modulation is smaller and it grows slower than those in Fig. 3(a). This indicates that the mechanism of the density modulation in Fig. 8 is not interaction-induced phase separation.
When the coupling strength is $\Omega_2= 2\pi \times 5600/15$ (= 373) Hz, 
the wavenumber of the density modulation is large and 
the amplitude is small [see Fig. \ref{fig: QuenchSim}(b)].

The quench experiment shown above is different from that proposed 
in Refs. \cite{Sabbatini_Phase_2011,Sabbatini_Causality_2012}. 
The initial state is assumed to be the ground state in the presence of Rabi coupling in Refs. \cite{Sabbatini_Phase_2011,Sabbatini_Causality_2012}, and does not undergo Rabi oscillation during the slow quench. In our experiment, however, the system does undergo Rabi oscillation during the slow quench. 
The phase separation instabilities and KZ scaling in such cases have yet to be studied.

\section{Discussion}\label{sec: Discussion}
The rf-induced interaction modulation in the long-lived states demonstrated here
is advantageous for the study of non-equilibrium miscibility-immiscibility dynamics over long durations.
Compared with optical Raman coupling \cite{Lin_Spin_2011},
rf coupling induces less heat due to the spontaneous emission
and imposes no practical limit on the coupling duration. 
Indeed, we confirmed experimentally that coupling does not shorten the lifetime of atoms.
Although the lifetime of atoms in our experiment is several seconds,
limited by other losses such as background gas collisions and three-body recombinations,
an experiment with $F=1$ states lasting more than 10 s will be possible \cite{De2014}.
In such a long-lived system, the relaxation dynamics toward equilibrium may be investigated.
Furthermore, the high controllability of the rf field allows us to engineer the coupling on demand
and to create arbitrary superposition of atomic states
including the ground dressed states \cite{Nicklas_Rabi_2011}, required for the proposed test of the quantum KZ theory \cite{Sabbatini_Phase_2011,Sabbatini_Causality_2012}.
The slow quench of the coupling with Rabi oscillation,
which we demonstrated here, will be another interesting nonequilibrium problem.

A drawback of the Rabi-induced interaction modulation is 
that the tuning range of the interaction is limited
without the aid of other methods, such as magnetic Feshbach resonance \cite{Papp_2008,Tojo_Controlling_2010}.
If the bare states are weakly miscible, the coupled states are weakly immiscible and vice versa.
This is the case for $^{87}$Rb atoms.
The use of other atoms with a larger interaction difference would be more suitable for observing deep phase separations.

\section{Conclusion}\label{sec: conclusion}
We have demonstrated interaction modulation in two pairs of the long-lived $F=1$ state in a $^{87}$Rb BEC.
We observed that the dynamics of the immiscible pair of $|1,-1 \rangle$ and $|1,0 \rangle$ states
and the miscible pair of $|1,-1 \rangle$ and $|1,+1 \rangle$ states
are altered by Rabi coupling.
The miscibility-to-immiscibility transition following a coupling quench in the $|1,-1\rangle$ and $|1,+1\rangle$ pair was also demonstrated.
We believe the rf-induced interaction modulation in the long-lived states 
is suitable for investigating unknown nonequilibrium dynamics.

\begin{acknowledgements}
We acknowledge support from MEXT/JSPS KAKENHI Grant Numbers JP15K05233, JP25103007,
JP17K05595, JP17K05596, JP16K05505.
\end{acknowledgements}


\end{document}